\documentclass{aa}

\usepackage[varg]{txfonts}
\usepackage{color}
\usepackage{graphicx}   
\usepackage{dcolumn}
\usepackage{bm}
\usepackage{dblfloatfix}
\usepackage{amsmath}    

\newcommand\simlt{\lower.5ex\hbox{$\; \buildrel < \over \sim \;$}}
\newcommand\simgt{\lower.5ex\hbox{$\; \buildrel > \over \sim \;$}}

\def \esp  {\epsilon^\prime_s}
\def \egp  {\epsilon^\prime_\gamma}
\def \eminp  {\epsilon^\prime_{min}}

\usepackage[colorlinks,citecolor=blue]{hyperref}

\begin{document}

\title{Particle-in-cell  simulations of pair discharges in a starved magnetosphere of a Kerr black hole}
\author{Amir Levinson$^{1,2}$ \and Beno\^it Cerutti$^{1}$}
\institute{Univ. Grenoble Alpes, CNRS, IPAG, 38000 Grenoble, France
\and Raymond and Beverly Sackler School of Physics \& Astronomy, Tel Aviv University, Tel Aviv 69978, Israel}

\date{Received / Accepted}

\abstract{We investigate the dynamics and emission of a starved magnetospheric region (gap) formed in the vicinity of a Kerr black hole horizon, using a new, fully general relativistic particle-in-cell code that implements Monte Carlo  methods to compute gamma-ray emission and pair production through the interaction of pairs and gamma rays with soft photons emitted by the accretion flow. It is found that when the Thomson length for collision with disk photons exceeds the gap width, screening of the gap occurs through low-amplitude, rapid plasma oscillations that produce self-sustained pair cascades, with quasi-stationary pair and gamma-ray spectra, and with a pair multiplicity that increases in proportion to the pair production opacity. The gamma-ray spectrum emitted from the gap peaks in the TeV band, with a total luminosity that constitutes a fraction of about $10^{-5}$ of the corresponding Blandford--Znajek power. This stage is preceded by a prompt discharge phase of duration $\sim r_g/c$, during which the potential energy initially stored in the gap is released as a flare of curvature TeV photons. We speculate that the TeV emission observed in M87 may be produced by pair discharges in a spark gap.}


\titlerunning{Pair discharges in a starved black hole magnetosphere}
\authorrunning{Amir Levinson \& Beno\^it Cerutti}

\maketitle

\section{Introduction}
The activation of  Blandford--Znajek (BZ) outflows requires continuous injection of plasma in the magnetospheric region enclosed between the inner and outer light cylinders. The origin of this plasma source is still an open issue. 
To fully screen out the magnetosphere, the plasma injection rate must be sufficiently high to maintain the density everywhere in the magnetosphere above the Goldreich--Julian (GJ) value.   If the plasma source cannot accommodate this requirement, charge starved regions will be created, potentially leading to self-sustained pair discharges.
A plausible plasma production mechanism in black hole (BH) engines is the annihilation of gamma rays that emanate from the accretion flow (e.g., \citealt{1977MNRAS.179..433B, 2000PhRvL..85..912L, 2007ApJ...671...85N, 2011ApJ...730..123L, 2016ApJ...818...50H, 2016ApJ...833..142H, 2017PhRvD..96l3006L}) or neutrino annihilation in the case of GRBs \citep{2014ApJ...796...26G}. Whether the pair density thereby produced is sufficiently high depends primarily on the luminosity of MeV photons emitted by the radiative inefficient accretion flow \citep{2011ApJ...730..123L, 2016ApJ...833..142H, 2018ApJ...852..112K} or by a putative corona in the case of sources accreting at a rate in excess of the critical ADAF rate (see, e.g., \citealt{2017PhRvD..96l3006L}).

It has been shown \citep{2011ApJ...730..123L, 2016ApJ...833..142H} that under conditions anticipated in many stellar and supermassive BH systems, the annihilation rate of disk photons is insufficient to maintain the charge density in the magnetosphere at the GJ value, giving rise to the formation  of spark gaps. It has been further pointed out \citep{2007ApJ...671...85N, Rie11, 2011ApJ...730..123L, 2016ApJ...818...50H, 2016ApJ...833..142H, 2017ApJ...845...40L, 2018ApJ...852..112K} that the gap activity may be imprinted in the high-energy  emission observed in these sources whereby  the variable TeV emission detected in M87 \citep{2003A&A...403L...1A, 2009Sci...325..444A} and IC310 \citep{2014Sci...346.1080A} was speculated to constitute examples of the signature  of magnetospheric plasma production on horizon scales \citep{2000PhRvL..85..912L, 2007ApJ...671...85N, 2011ApJ...730..123L, 2016ApJ...818...50H}.

In an attempt to compute the emission properties of an active gap, a fully general relativistic (GR) steady gap model has recently been developed  \citep{2016ApJ...818...50H, 2016ApJ...833..142H, 2017PhRvD..96l3006L} that incorporates curvature emission, inverse Compton scattering, and pair creation via the interaction of gamma rays produced in the gap with the external photons emanating from the accretion disk. However, as argued in \citet{2017PhRvD..96l3006L}, steady-state solutions are restricted to a narrow range of conditions that may not apply to most systems. Furthermore, it is unclear whether the steady gap solutions obtained in the works cited above are stable in the first place.

In this paper we explore the dynamics of a local, 1D gap using a new particle-in-cell (PIC) code, developed  particularly for this purpose, that implements Monte Carlo  methods to compute gamma-ray emission and pair production through the interaction of pairs and gamma rays with an external radiation field. The details are described in \S \ref{sec:analysis} and \S \ref{sec:grpic} below. We find (\S \ref{sec:results}) that after a prompt discharge phase that screens out the gap and produces a strong gamma-ray flare, the system relaxes to a state of self-sustained, rapid plasma oscillations that is independent of the initial conditions.   The amplitude of the oscillating electric field in the relaxed state is regulated by pair production, such that the average pair creation rate inside the simulation box equals the rate at which pairs escape through the boundaries. The pair and photon spectra are quasi-stationary during this state. We find that the saturated pair multiplicity increases roughly linearly with the opacity contributed by the ambient radiation field, while the average pair and photon energies decrease as the opacity increases. The gamma-ray luminosity radiated by the accelerating pairs following the prompt phase (after relaxation of the system) constitutes a fraction of about $10^{-5}$ of the corresponding BZ power, weakly dependent on the pair production opacity and other model parameters.  As pointed out in \S \ref{sec:M87}, this is in rough agreement with the TeV observations of M87.

Our approach is similar to that used by \citet{2010MNRAS.408.2092T} and \citet{2013MNRAS.429...20T}, with the following  differences: it is fully GR, it has no external plasma source (the neutron star), and it includes inverse Compton scattering and pair production via interactions with an external radiation field in addition to curvature emission.

\section{\label{sec:analysis}Oscillating gap model}
We study the dynamics of a local, 1D gap using fully general relativistic PIC simulations.  
We assume that radiation emanating from the inner regions of the accretion flow provides the dominant source of opacity for pair production and inverse Compton scattering. The intensity of this radiation is given as an input for the computations of the gap dynamics. 
Gamma rays generated via inverse Compton scattering of the ambient radiation by the pairs accelerated in the gap are treated as a neutral species in our PIC scheme (in addition to electrons and positrons). Pair production occurs through the interaction of the gamma rays thereby produced with the external photons emitted by the accretion flow.  The various radiation processes are computed using a novel Monte Carlo method developed for this purpose (see Appendix \ref{app:MC} for details). In addition, we include curvature losses in the equations of motions and provide a rough estimate for the luminosity of curvature emission, as explained below. However, for numerical reasons we do not add the curvature photons to the pull of gamma rays, and therefore do not account for gamma ray production by the curvature photons.  As shown below, for cases of interest curvature emission is only important during the initial discharge of the gap. 

We implicitly assume that the gap constitutes a small disturbance in the magnetosphere, in the sense that its activity does not significantly affect the global structure, and in particular the magnetic field geometry and the angular velocity of magnetic surfaces $\Omega$. The coupling between the gap and the global magnetosphere enters through the global electric current flowing in the magnetosphere, treated as a free  input parameter, as in the steady-state models \citep{2016ApJ...818...50H, 2017PhRvD..96l3006L}.
For simplicity we adopt a split monopole geometry, defined by $A_\varphi=C(1-\cos\theta)$.
For this choice $F_{r\varphi}=0$, and from the ideal MHD condition we obtain $F_{rt}=-\Omega F_{r\varphi}=0$ for the radial electric field outside the gap, in the ideal MHD sections of the magnetosphere. This effectively ignores any MHD waves that might be generated by the gap
cycle and propagate throughout the force-free magnetosphere. The gap extends along a poloidal magnetic surface, characterized by an inclination angle $\theta$. Inside the gap $F_{rt}\ne 0$ by virtue of the pair creating oscillations.  This is the only wave field that appears in the dynamical equations derived below, hence the gap activity is restricted to longitudinal plasma oscillations in this model.  
Despite this restriction, this model captures the main features of plasma production and consequent emission. We consider it as a preliminary stage in our quest for the development of a 2D code that computes the global structure and dynamics of the magnetosphere. 

\subsection{Background geometry and choice of coordinates}
The background spacetime is described by the Kerr metric, here given in Boyer--Lindquist coordinates with the notation
\begin{equation}
ds^2= -\alpha^2dt^2 + g_{\varphi\varphi}(d\varphi-\omega dt)^2 + g_{rr}dr^2 + 
g_{\theta\theta}d\theta^2,
\end{equation}
where 
\begin{eqnarray}
 \alpha^2&=&\frac{\Sigma\Delta}{A};\quad \omega=\frac{2ar_g r}{A};\quad g_{rr}=\frac{\Sigma}{\Delta};\\
&g_{\theta\theta}&=\Sigma; \quad g_{\varphi\varphi}=\frac{A}{\Sigma}\sin^2\theta, \nonumber
\end{eqnarray}
with $\Delta=r^2+a^2-2r_g r$,  $\Sigma=r^2+a^2\cos^2\theta$,
$A=(r^2+a^2)^2-a^2\Delta\sin^2\theta$, and $r_g=GM/c^2 = 1.5\times10^{14}\, M_9$ cm denotes the gravitational radius, and $M=10^9 M_\odot$ the BH mass. The parameter $a=J/M$ represents the specific angular momentum.  The determinant of the matrix $g_{\mu\nu}$ is given by $\sqrt{-g}=\Sigma\sin\theta$. The angular velocity of the black hole is defined as $\omega_H=\omega(r=r_H)=\tilde{a}/2r_H$, where $\tilde{a}=a/r_g$ denotes the dimensionless spin parameter, and $r_H=r_g+\sqrt{r_g^2-a^2}$ is the radius of the horizon. Hereafter, all lengths are measured in units of $r_g$ and time in units of $t_g=r_g/c$, so we set $c=r_g=1$, unless explicitly stated otherwise.

To avoid the singularity on the horizon, we find it convenient to transform to the tortoise coordinate $\xi$, defined by $d\xi= dr/\Delta$ (in full units $d\xi=r_g^2 dr/\Delta$). It is related to $r$ through
\begin{equation}
\xi(r)=\frac{1}{r_+-r_-}\ln\left( \frac{r-r_+}{r-r_-} \right),
\end{equation}
with $r_\pm=1\pm \sqrt{1-\tilde{a}^2}$.  We note  that $\xi\rightarrow-\infty$ as $r\rightarrow r_H=r_+$, and $\xi \rightarrow 0$ as $r\rightarrow\infty$. We prefer to use this version of the tortoise coordinate for the following reasons: (i) it pushes the horizon to  $-\infty$, thereby allowing us to conveniently choose the inner gap boundary as close to the horizon as needed; (ii) when using quantities measured in the frame of a zero angular momentum observer (ZAMO), the form of the equations in these coordinates is very similar to that in flat spacetime. This renders interpretation of the results more intuitive; and, most importantly,  (iii)  it does not mix the radial coordinate with the time and azimuthal coordinates, as it does in the case of Kerr--Schild coordinates, for example. This greatly simplifies the Monte Carlo computations of Compton scattering and pair production.   When using  Kerr--Schild coordinates additional transformations are needed in every computation step, because the emissivity and absorption coefficient are naturally defined in the ZAMO frame. This is an unnecessary complication that is avoided by our choice of the tortoise coordinates.

For a grid cell $\delta r$ on a magnetic surface having an inclination angle $\theta$, 
a volume element can be defined as
\begin{align}
\delta V & =2\pi \int_{r-\delta r/2}^{r+\delta r/2}\int_{\theta}^{\theta+\delta\theta}  \sqrt{-g}d\theta^\prime dr^\prime\\ \nonumber
&=2\pi \int_{r-\delta r/2}^{r+\delta r/2}dr^\prime\,\int_{\mu}^{\mu+\delta\mu} (r^2+a^2\mu^{\prime 2})d\mu^\prime\\ \nonumber
&= 2\pi \int_{\xi-\delta \xi/2}^{\xi+\delta \xi/2}\Delta d\xi^\prime\,\int_{\mu}^{\mu+\delta\mu} (r^2+a^2\mu^{\prime 2})d\mu^\prime\equiv \Delta\, \delta V_\xi.
\end{align}
Here, $\delta V_\xi$ defines the volume element with respect to the coordinate $\xi$, and is finite on the horizon.  The average number density (of either pairs or gamma rays) within a grid cell, as measured by a distant observer, is given by
\begin{equation}
n=\delta N/\delta V = \Delta^{-1}\delta N/\delta V_\xi \equiv \Delta^{-1} n_\xi,
\end{equation}
where $\delta N$ denotes the occupation number  of the grid cell.  The quantity $n_\xi = \Delta n$ is finite on the horizon, and will be used hereafter to describe the density of the different plasma constituents. 

\subsection{Basic equations}
\subsubsection{Electrodynamics}
As explained above, the only wave field in our model is the radial component of the electric field, $F_{rt}$ (see  Appendix \ref{sec:EM} for details). Measured in the ZAMO frame it reads $E_r=\sqrt{A} F_{rt}/\Sigma$. Its dynamics is governed by Equation (\ref{app:dtEr}), here expressed in terms of the electric flux per steradian, $\sqrt{A}E_r$, as
\begin{equation}
\partial_t(\sqrt{A} E_r) =- 4\pi(\Sigma j^r-J_0),\label{dEdt}
\end{equation}
where $J_0=\Sigma j^r_0$ is the global magnetospheric current, defined explicitly in Equation (\ref{J_0}).  This current is conserved along magnetic surfaces, and serves as an input parameter to the dynamic gap model. Gauss's law  further yields
\begin{equation}
\partial_\xi(\sqrt{A} E_r)=4\pi \Delta\, \Sigma (j^{t}-\rho_{GJ}),\label{Poiss}
\end{equation}
where 
\begin{equation}
\rho_{GJ}=\frac{B_H\sqrt{A_H}}{4\pi \sqrt{-g}} \left[\frac{\sin^2\theta}{\alpha^2}(\omega-\Omega)\right]_{,\theta}
\label{rho_GJ}
\end{equation}
is the  GJ density, given explicitly in Equation (\ref{app:rho_GJ_exp}), $B_H$ is the strength of the magnetic field on the horizon, and $A_H\equiv A(r_H)$.  
We note that charge conservation readily implies that Eq. (\ref{Poiss}) is conserved in time.   That is, if the initial state satisfies this equation, and the system then evolves in time according to Eq. (\ref{dEdt}), it is guaranteed that  Eq. (\ref{Poiss}) will be satisfied at any given time. Thus, this equation is only used once at the beginning of each run to determine the initial state of the electric field, $E_r(r,t=0)$, for a given choice of initial conditions.  

\subsubsection{Particle motion}

The  plasma in the gap consists of electrons and positrons. Let $u_i^\mu$ denote the four-velocity of the ith particle and $q_i$ its electric charge, where $q_i=-e$ for electrons and $q_i= +e$ for positrons.  
The equation of motion for the ith particle can then be expressed as
\begin{eqnarray}
\frac{du_i^\mu}{d\tau_i}=-\Gamma^\mu_{\, \, \,  \alpha\beta}u_i^\alpha u_i^\beta+\frac{q_i}{m_e}F^\mu_{\, \, \, \alpha}u_i^\alpha+s_i^\mu ,
\label{EOM_h}
\end{eqnarray}
subject to the normalization $u_i^\mu u_{i\,\mu}=-1$, where $s_i^\mu$ is a source term associated with curvature losses, satisfying $u_{i \mu} s_i^\mu=0$
, and 
$d\tau_i$ is the corresponding proper time interval. It is related to the  time measured by a distant observer through $dt=u^t_i d\tau_i$. For the lowered index components, $u_{i\mu}=g_{\mu\nu}u_i^\nu$, we likewise have
\begin{eqnarray}
\frac{du_{i\,\mu}}{d\tau_i}=\Gamma_{\alpha\, \mu\, \beta}u_i^\alpha u_i^\beta+\frac{q_i}{m_e}F_{\mu\alpha}u_i^\alpha+s_{i\, \mu}.
\label{EOM_low}
\end{eqnarray}

Poloidal motion is restricted to the radial direction in our 1D model, thus $u_i^\theta=0$.  Furthermore, since $\partial_\varphi$ is a Killing vector it readily follows that $\Gamma_{\alpha\, \varphi\, \beta}u_i^\alpha u_i^\beta=0$, and since $F_{\varphi t}=F_{\varphi r}=0$ we obtain 
$F_{\varphi \alpha}u_i^\alpha=F_{\varphi \theta}u_i^\theta=0$.   As the curvature drag term, $s_{i\varphi}$, is proportional to $u_\varphi$,
it is evident from Equation (\ref{EOM_low}) that particles tend to a state of zero angular momentum, hence we set $u_\varphi=s_{i\varphi}=0$,  
for which we obtain $s^\varphi_i=\omega s_i^t$ and $s_{it}=-\alpha^2 s_i^t$.
We find it convenient to use the four-velocity components measured by a ZAMO, here denoted by $u_i=\sqrt{g_{rr}}u_i^r$, the particle Lorentz factor $\gamma_i=\alpha u^t_i$, and the three-velocity $v_i=u_i/\gamma_i$. The normalization condition yields $\gamma_i^2=1+u_i^2$.   Using Equations (\ref{EOM_h}) and (\ref{EOM_low}) and the relations $2u_idu_i/d\tau_i = u_{ir}du^r_i/d\tau_i+ u^r_idu_{ri}/d\tau_i$, $d\tau_i=dt/u^t_i$, and $u^r_i s_{ir}+u^t_i s_{it}=0$, we arrive at 
\begin{eqnarray}
\frac{du_{i}}{dt}&=&\sqrt{g^{rr}} \left[-\gamma_i\partial_r (\alpha) +\frac{q_i}{m_e}F_{r t}\right]-\frac{s_{it}}{u_{i}} \nonumber \\ 
&=&-\sqrt{g^{rr}} \gamma_i\partial_r (\alpha)+\alpha\left(\frac{q_i}{m_e}E_r -\frac{P_{cur}(\gamma_i)}{m_e v_i}\right),
\label{EOM_pold}
\end{eqnarray}
where $-s^t/u^t=P_{cur}(\gamma_i)$ is the curvature power emitted by the particle, as measured by a ZAMO (Equation (\ref{P_cur}) below), and
\begin{equation}
\partial_r \left(\alpha\right)=\frac{\alpha }{A}\left(\frac{2r^2\tilde{a}^2\sin^2\theta}{\Sigma}+\frac{r^4-\tilde{a}^4}{\Delta}\right).
\end{equation}
The particle trajectory is computed using
\begin{equation}
\frac{d\xi_i}{dt}  =\frac{1}{\Delta} \frac{dr_i}{dt}=\frac{1}{\Delta}\frac{u_i^r}{u_i^t}=\frac{v_i}{\sqrt{A}}.
\end{equation}
In terms of $v_i$ the electric current density in Equation (\ref{dEdt}) can be expressed as
\begin{equation}
j^r=\frac{1}{\delta V}\sum_{i\in\delta V} q_i\frac{u_i^r}{u_i^t}=\frac{1}{\sqrt{A}\delta V_\xi}\sum_{i\in \delta V} q_i\, v_i.
\label{eqjr}
\end{equation}
Likewise, the charge density in Equation (\ref{Poiss}) satisfies
\begin{equation}
\Delta\, j^t=\frac{1}{\delta V_\xi}\sum_{i\in\delta V}q_i.
\end{equation}

\subsubsection{Radiation}
Gamma rays are treated as a neutral species in our scheme.  Let $\tilde{p}^\mu_k$ denote the momentum of the kth photon, as measured by a ZAMO. It is related to the coordinate momentum $p^\mu_k$ though $\tilde{p}_k^t=\alpha p_k^t$, $\tilde{p}_k^r=\sqrt{g_{rr}}\, p_k^r$,
$\tilde{p}_k^\theta=\sqrt{g_{\theta\theta}}\, p_k^\theta$, and $\tilde{p}_k^\varphi=p_{k\, \varphi}/\sqrt{g_{\varphi\varphi}}$. 
At the energies of interest the gamma rays are highly beamed along the direction of motion of the emitting particles by virtue of momentum conservation, thus to a very good approximation we can set $\tilde{p}_k^\theta=\tilde{p}_k^\varphi=0$. The normalization condition, $p_{k\mu}p_k^\mu=0$, gives $\tilde{p}_k^t=|\tilde{p}_k^r|$, as required by the fact that photons propagate at the speed of light in the ZAMO frame. The null geodesic equations reduce to 
\begin{equation}
\frac{d\tilde{p}_k^r}{dt}=-\sqrt{g^{rr}} \tilde{p}_k^t \partial_r \left(\alpha\right),
\label{eqphotons}
\end{equation}
and the photon trajectory equation to 
\begin{equation}
\frac{d\xi_k}{dt}  =\frac{1}{\sqrt{A} }\frac{\tilde{p}_k^r}{\tilde{p}_k^t}.
\end{equation}

We suppose that the gap is exposed to the emission of soft photons by the accretion flow 
from a putative source of size $R_s=\tilde{R}_s r_g$ and luminosity $L_s=l_s L_{Edd}$.  
For simplicity, we assume that the intensity of the seed radiation inside the simulation box is isotropic, constant in time, and homogeneous, with a power law spectrum
\begin{equation}
I_s(x^\mu,\epsilon_s,{\bf \Omega}_s)=I_0(\epsilon_s/\epsilon_{s,min})^{-p},  \quad \epsilon_{s,min}<\epsilon_s<\epsilon_{s,max},
\label{Intensity-1}
\end{equation}
with $p>1$, where $\epsilon_s$ is the photon energy in $m_ec^2$ units, as measured by a ZAMO.  
The assumption that $I_s$ is isotropic is reasonable, except perhaps very near the horizon, 
since the size $R_s$ of the radiation source is typically larger than the gap dimensions.
The total number density of photons is given approximately by  $n_s= 4\pi \int (I_s/hc\epsilon_s)d\epsilon_s \simeq 4\pi I_0/hc$,
and can be used to define the fiducial optical depth
\begin{equation}
\tau_0=4\pi r_g \sigma_T I_0/hc\simeq \frac{4 m_p}{m_e}\frac{l_s}{\tilde{R}_s^2\epsilon_{min}}.
\label{tau_0}
\end{equation}
For $\tilde{R}_s=10$ we have $l_s\simeq 10^{-2}\epsilon_{min}\tau_0$.

The interaction of pairs and gamma rays with the external radiation field is computed using a Monte Carlo approach.  The details are given in Appendix \ref{app:MC}. In short,  for every particle we compute the optical depth traversed by the particle between two consecutive  time steps, $t$ and $t+\delta t$, along its trajectory $r(t)$, using the full Klein--Nishina opacity $\kappa_c$, given in Equation (\ref{k_c}):
\begin{equation}
\delta\tau_{sc}=\int^{r(t+\delta t)}_{r(t)} \kappa_c \sqrt{g_{rr}} dr=  \int_{\xi(t)}^{\xi(t+\delta t)}\kappa_c \sqrt{\Sigma \Delta} d\xi.
\label{Dtau_c}
\end{equation}
We then randomly draw a probability for a scattering event: $0\le p_{sc}\le 1$.  A scattering event occurs if $p_{sc}<1-e^{-\delta \tau_{sc}}$. The process is repeated at every time step for all particles in the simulation box. If  a scattering event occurs, we transform to the
rest frame of the particle and draw the energy and direction of the scattered gamma ray, as explained in Appendix \ref{app:MC}, and then transform back to the ZAMO frame and compute the new energy of the particle.   The newly created gamma ray is added to the pull. 

Pair production via the interaction of gamma rays with the external radiation field is computed in a similar manner. We use Equation (\ref{Dtau_c}) with $\kappa_c$ replaced by the pair production opacity $\kappa_{pp}$ (Eq. (\ref{k_PP})) to calculate the optical depth $\delta\tau_\pm$ traversed by a gamma ray between two consecutive time steps. Pair production occurs if  $p_{\pm}<1-e^{-\delta \tau_\pm}$, upon drawing a probability $0\le p_{\pm}\le 1$.  
The process is repeated at every time step for all gamma rays in the simulation box.  The pull of gamma rays and pairs is updated correspondingly. To simplify the analysis we suppose that in each pair production event the newly created electron and positron have identical energies.   This is a good approximation near the threshold, where the cross section is at maximum (e.g., \citealt{1995ApJ...441...79B}). 

The curvature power emitted by an electron (positron) of energy $\gamma$ in the ZAMO frame is given by
\begin{equation}
P_{cur}(\gamma)=\frac{2}{3}\frac{e^2  \gamma^4 v^4}{R_c^2},
\label{P_cur}
\end{equation}
where $R_c$ is the curvature radius of the particle trajectory.  For the computations presented below we adopt $R_c=r_g$.
The characteristic energy of a curvature photon (measured in $m_ec^2$ units) is
\begin{equation}
\epsilon_{c}=\frac{2\pi \lambda_c}{R_c}\gamma^3=10^{-24}M_9^{-1}\gamma^3,
\label{eps_c}
\end{equation}
where $\lambda_c=\hbar/m_ec$ is the Compton wavelength of the electron.  
The maximum Lorentz factor of the pairs is limited by back reaction. From Equation (\ref{EOM_pold}) we find 
\begin{equation}
\gamma<\gamma_{max}=\left(\frac{E_0 R_c^2}{e}\right)^{1/4}\simeq 1.7\times10^{10}(\eta B_3)^{1/4}M_9^{1/2},
\label{g_max}
\end{equation}
where $E_0=\eta B_H=\eta 10^3B_3$ G is the maximum strength of the electric field in the gap (see middle left panel in Fig. \ref{fig:ex1}). This maximum value is delineated by the vertical dashed line in Fig. \ref{fig:ex1b}. As will be shown, in the initial discharge pairs indeed accelerate to this Lorentz factor; however, after the relaxation of the system the Lorentz factor is essentially limited by the oscillations of the electric field rather than curvature losses, provided $\tau_0>1$. In a transparent gap, $\tau_0<1$, we expect domination of curvature radiation during the entire evolution of the system.

The number of curvature photons emitted by an electron (positron) of Lorentz factor $\gamma$ over a dynamical time $t_g=r_g/c$ is
\begin{equation}
N_{cur}=P_{cur}t_g/\epsilon_c=\frac{e^2}{3\pi \hbar c}\gamma=7\times10^{-4}\gamma.
\end{equation}
It varies between $10^7$ during the initial discharge (where $\gamma\simeq\gamma_{max}$) to about $10^5$ during the relaxed state. Thus, without proper resampling it is practically infeasible to track these photons in our PIC scheme. We note that once $\gamma $ drops below $10^9$ the characteristic energy of curvature photons becomes $\epsilon_c<10^3$, and their contribution to pair creation and to the gamma-ray power emitted from the gap can be neglected. However, curvature emission dominates the released power (and likely pair creation) during the initial discharge. In order to account for this power, yet avoiding tremendous numerical complications, the net curvature luminosity measured by a distant observer is computed, at any given time step, by summing up the contributions of all particles in the simulation box, accounting properly for redshift effects
\begin{equation}
L_{cur}=\frac{1}{2}\sum_i \alpha_i^2 P_{cur}(\gamma_i),
\label{L_curv}
\end{equation}
with $\alpha_i\equiv \alpha(r_i)$ denoting the lapse function of the ith particle, currently located  at radius $r_i$.   This prescription lacks proper treatment of time travel effects, which are  particularly important near the horizon.  It merely provides a rough estimate of the contribution of curvature emission to the emitted gamma-ray luminosity.   

\subsection{Input parameters}

The 1D gap model is characterized by the following input parameters: the black hole mass ($M=10^9 M_9 ~ M_\odot)$, the BH spin parameter ($\tilde{a}$), the angular velocity ($\Omega$) of the magnetic surface along which the gap lies, the strength of the magnetic field on the horizon ($B_H=10^3 B_3$ G), the inclination angle of magnetic surface ($\theta$), the minimum energy and slope of the target spectrum ($\epsilon_{s,min}$ and $p$) (Eq. \ref{Intensity-1}), the global electric current ($J_0$) defined in Equation (\ref{J_0}), and the fiducial optical depth ($\tau_0$) defined in Equation (\ref{tau_0}). The results presented below were computed using the following canonical choice of parameters ($M_9=1$, $B_3=2\pi$, $\theta=30^\circ$, $\tilde{a}=0.9$, $\epsilon_{s,min}=10^{-8}$, $p=2$), which correspond to supermassive BHs accreting in the RIAF regime. For this choice of parameters we explore how the behavior of the solutions depends on $J_0$ and $\tau_0$. 

\section{\label{sec:grpic} Implementation of the 1D GRPIC code}

\subsection{Numerical methods}

The overall structure of the code is based on the 1D version of the {\tt ZELTRON} code \citep{2013ApJ...770..147C}, a highly parallelized (special) relativistic electromagnetic PIC code, in addition to which GR corrections and the Monte Carlo scheme introduced in the previous section were implemented for this study. The code solves the equations of motions and Maxwell's equation using an explicit second-order finite-difference scheme. 

The electric field is evolved in time using Eq.~(\ref{dEdt}), such that
\begin{equation}
\sqrt{A}E^{n+1}_{\rm r, i\xi}=\sqrt{A}E^{n}_{\rm r, i\xi}-4\pi\delta t\left(\Sigma j^{r,n+1/2}_{i\xi}-J_0\right),
\end{equation}
where $\delta t=t^{n+1}-t^{n}=t^{n+1/2}-t^{n-1/2}$ is the time step. The electric field is defined at full time steps $n$ and $n+1$, while the current is defined at half time steps $n+1/2$. This offset in time ensures a second order accuracy of the scheme. Both $E_r$ and $j^r$ are defined at the nodes of the grid given by the index $i\xi$. Currents from individual charged particles are deposited on the grid using Eq.~(\ref{eqjr}) where the three-velocities are known at the half time step $n+1/2$.

In the 1D limit, the equations of motions are also  straightforward to integrate even though more steps are needed than for the field. Starting with the equation of motion of the photons (Eq.~\ref{eqphotons}), a time-centered scheme gives
\begin{equation}
\frac{\tilde{p}_k^{r,n+1/2}-\tilde{p}_k^{r,n-1/2}}{\delta t}=-\sqrt{g^{rr}}\left|\tilde{p}_k^{r,n}\right|\partial_r\left(\alpha\right).
\end{equation}
Assuming that $\tilde{p}_k^{r,n}=(\tilde{p}_k^{r,n+1/2}+\tilde{p}_k^{r,n-1/2})/2$, we obtain
\begin{equation}
\tilde{p}_k^{r,n+1/2}=\left(\frac{1\mp\sqrt{g^{rr}}\partial_r\left(\alpha\right)\delta t/2}{1\pm\sqrt{g^{rr}}\partial_r\left(\alpha\right)\delta t/2}\right)\tilde{p}_k^{r,n-1/2},~{\rm sgn}\left(\tilde{p}_k^{r}\right)=\pm 1.
\end{equation}
The photon position at time $t^{n+1}$ is then given by
\begin{equation}
\xi^{n+1}_k=\xi^{n}_k+\frac{\delta t}{\sqrt{A}}\frac{\tilde{p}_k^{r,n+1/2}}{\left|\tilde{p}_k^{r,n+1/2}\right|}.
\end{equation}
For the charged particles, we have (see Eq.~\ref{EOM_pold})
\begin{equation}
\frac{u^{n+1/2}_i-u^{n-1/2}_i}{\delta t} = -\gamma^{n}_i\sqrt{g^{rr}}\partial_r\left(\alpha\right)+\alpha\frac{q_i}{m_{e}}E^{n}_r-\alpha\frac{P_{cur}\left(\gamma^n_i\right)}{m_{e}v^{n}_i}.\\
\end{equation}
We break up this equation into three components corresponding to each term on the right-hand side
\begin{eqnarray}
\frac{u^{n+1/2}_{i,g}-u^{n-1/2}_{i,g}}{\delta t} &=& -\gamma^{n}_i\sqrt{g^{rr}}\partial_r\left(\alpha\right),\\
\frac{u^{n+1/2}_{i,L}-u^{n-1/2}_{i,L}}{\delta t} &=& \alpha\frac{q_i}{m_{e}}E^{n}_r,\label{equL}\\
\frac{u^{n+1/2}_{i,R}-u^{n-1/2}_{i,R}}{\delta t} &=& -\alpha\frac{P_{cur}\left(\gamma^n_i\right)}{m_{e}v^{n}_i}.
\end{eqnarray}
Assuming that $u^{n-1/2}_{i,g}=u^{n-1/2}_{i,L}=u^{n-1/2}_{i,R}=u^{n-1/2}_i$ and adding all three equations together yields
\begin{equation}
u^{n+1/2}_i=u_{i,L}^{n+1/2}+\left(u_{i,g}^{n+1/2}-u_i^{n-1/2}\right)+\left(u_{i,R}^{n+1/2}-u_i^{n-1/2}\right).
\label{eqmotion}
\end{equation}
The next step is to estimate the particle Lorentz factor and three-velocity at time step $t^n$. To do this, we first compute $u_{i,L}^{n+1/2}$ using Eq.~(\ref{equL}) where $E^{n}_r$ is already known at the nodes of the grid and linearly interpolated at the particle position. Using the same trick as for the photons, {i.e.}, $u_{i,L}^{n}=(u_{i,L}^{n+1/2}+u_{i,L}^{n-1/2})/2$, we compute
\begin{equation}
\gamma^{n}_{i,L}=\sqrt{1+\left(u^n_{i,L}\right)^2},~v^n_{i,L}=\frac{u_{i,L}^n}{\gamma_{i,L}^n}.
\end{equation}
The last step is to inject these estimated values into Eq.~(\ref{eqmotion}), such that
\begin{equation}
u^{n+1/2}_i\approx u^{n+1/2}_{i,L}-\delta t \left(\gamma_{i,L}^{n}\sqrt{g^{rr}}\partial_r\left(\alpha\right)+\alpha\frac{P_{cur}\left(\gamma_{i,L}^n\right)}{m_{e}v_{i,L}^{n}}\right).
\end{equation}
The charged particle positions are updated in the same way as for photons,
\begin{equation}
\xi^{n+1}_i=\xi^{n}_i+\frac{\delta t}{\sqrt{A}} v^{n+1/2}_i.
\end{equation}

\subsection{Numerical setup}

The spatial grid is fixed in time and uniform in $\xi$, ranging from $\xi_{\rm min}=-3$ ($r_{\rm min}\approx 1.5$) to $\xi_{\rm max}=-0.3$ ($r_{\rm max}\approx 4.3$). Therefore, the grid is highly non-uniform in radius, refined near the BH horizon and sparse in the outer regions of the box. The time step is set by the Courant-Friedrichs-Lewy condition defined at the inner boundary, i.e.,
\begin{equation}
\delta t\leq \delta t_{\rm CFL}\equiv\left(r^2_{\rm min}+\tilde{a}^2-2r_{\rm min}\right)\delta\xi.
\end{equation}
The simulation box is initially filled with a monoenergetic beam of gamma-ray photons uniformly distributed along the $\xi$-grid for reasons explained below, but it is empty of charged particles. The initial (vacuum) electric field is obtained by numerically integrating Eq.~(\ref{Poiss}) with $j^t=0$. The solution is shown in the middle-left panel in figure~\ref{fig:ex1}. Gauss's law is integrated only once at the beginning of the simulation. We have checked that it is well satisfied throughout the simulation by virtue of charge conservation.

The choice of boundary conditions for the fields are trivial in this problem because Amp\`ere's law given in Eq.~(\ref{dEdt}) does not involve any spatial derivative in 1D. The electric field is free to evolve throughout the box. The particles, whether  charged or neutral, are simply deleted from the memory as soon as they cross the boundaries to mimic an open boundary on both sides. Therefore, we assume that no plasma injection from outside is permitted.

Spatial and temporal resolutions are harder quantities to define in this setup because they essentially depend on the energy and the density of particles, which are the unknowns that we are trying to measure here. Thus, resolution and numerical convergence can be checked a posteriori only. More specifically, it is crucial to resolve the collisionless plasma skin depth, $l_p$, otherwise the plasma will heat up artificially until it is resolved by the grid. It is instructive to give an estimate of the skin depth in terms of the model parameters, the pair multiplicity, $\kappa=n_\pm/n_{GJ}$, where $n_\pm$ denotes the pair density, and the mean Lorentz factor of pairs $\langle\gamma\rangle$. Recalling that the plasma frequency is $\omega_p= \sqrt{4\pi e^2 n_{\pm}/m_e\langle\gamma\rangle}$ and $e n_{GJ}=\Omega B/2\pi c$, and adopting $\Omega=\omega_H/2$, we obtain
\begin{equation}
l_p\equiv \frac{c}{\omega_p}=\sqrt{\frac{m_ec^3\langle\gamma\rangle}{\omega_H B e \kappa}}, 
\end{equation}
and
\begin{equation}
\frac{r_g}{l_p} \simeq 10^7\sqrt{\frac{\kappa M_9 B_3}{2\langle\gamma\rangle}}.
\end{equation}
The  number of cells needed to resolve the skin depth in a simulation box of size $h$ is approximately $h/l_p=(h/r_g)(r_g/l_p)$. In the cases studied below we find $\kappa/\langle\gamma\rangle$ in the range $10^{-9} - 10^{-7}$ for a range of opacities $\tau_0=1$ to $\tau_0=10$. For the size of our simulation box, $h/r_g\approx 3$, at least $10^4$ cells are needed to resolve the skin depth for $\tau_0=10$. Simulations with $\tau_0<10$ showed good convergence for a total $N_{\xi}=16384$ cells. For $\tau_0=10$, we had to run with up to $N_{\xi}=65536$ cells to see convergence. For the results described below, the initial beam of gamma rays is modelled with five particles per cell. At the end of the simulation, the average number of particles per cell varies from $2$ ($\tau_0=1$) to $10$ ($\tau_0=10$) for the pairs and from $2$ to $50$ for the photons. We have also checked the good convergence of our results with respect to the initial number of injected gamma-ray photons per cell. 

\section{\label{sec:results}Results}

\subsection{Overall evolution}

\begin{figure*}[h]
\centering
\includegraphics[width=8cm]{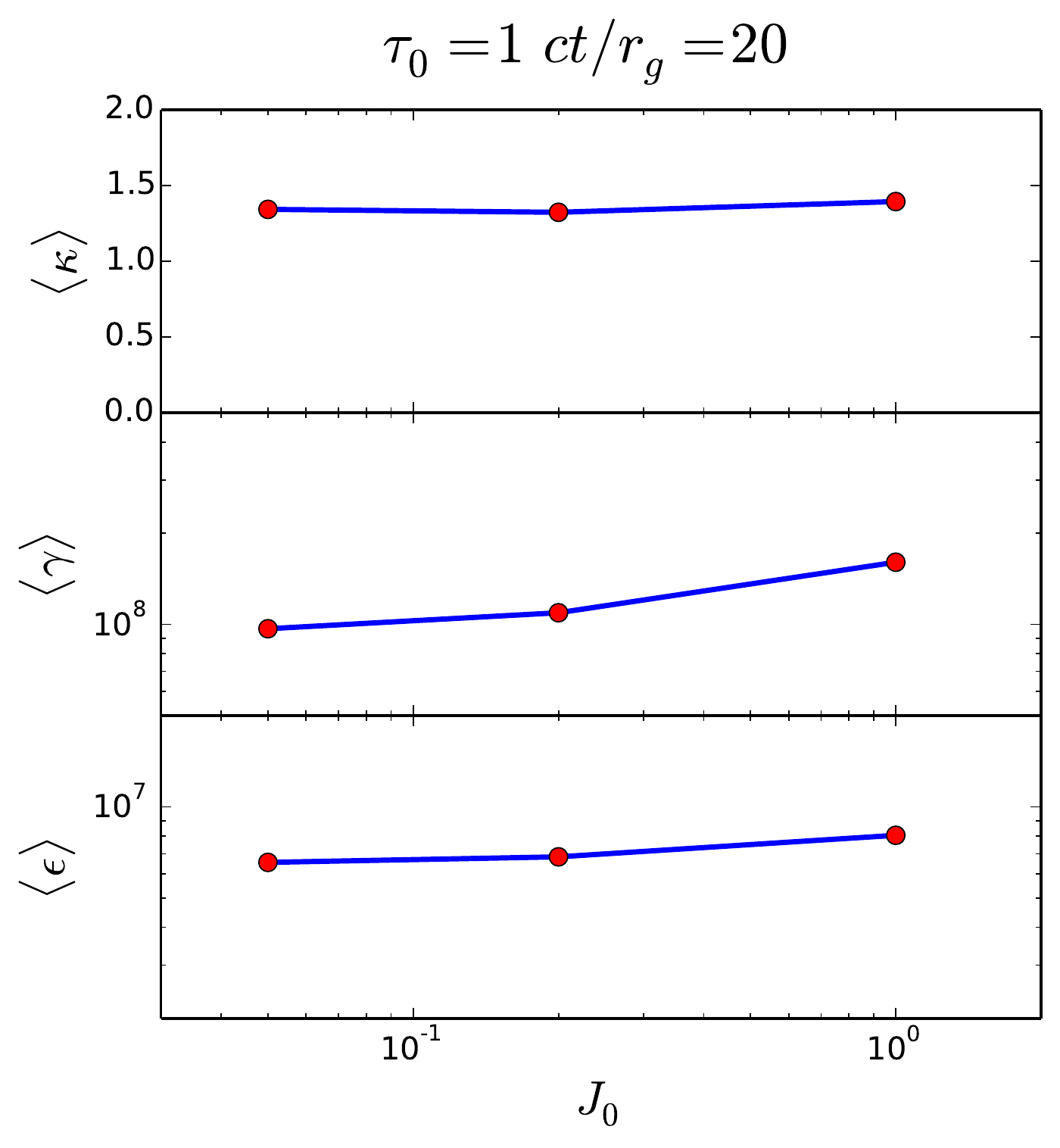} \includegraphics[width=8cm]{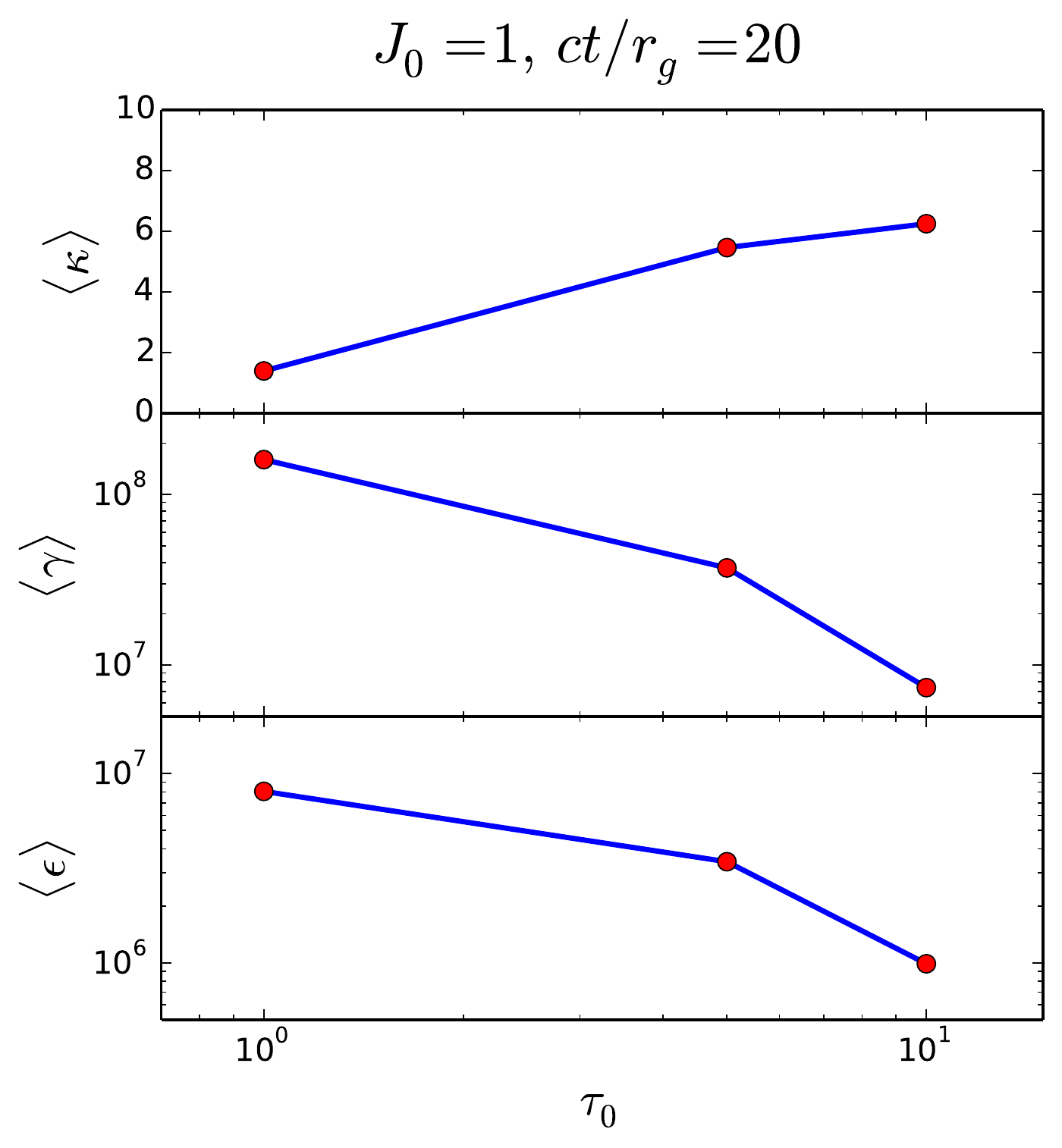} 
\caption{Dependence of the pair multiplicity $\langle\kappa\rangle$, pair energy $\langle\gamma\rangle$, and gamma-ray energy $\langle\epsilon\rangle$ averaged over the simulation domain as functions of the input parameters $\tau_0$ (right) and $J_0$ (left).}   
\label{fig:j0}
\end{figure*}

We  ran simulations for a grid of models characterized by different values of the parameters $\tau_0$ and $J_0$. Quite generally, we find  that the evolution of the system depends on the value of $\tau_0$, but is practically independent of $J_0$ (Fig. \ref{fig:j0}).
The prime role of $J_0$ is to fix the time average value of the oscillating current.
In all cases explored we find an initial discharge of the gap that produces a gamma-ray flare, dominated by curvature losses, with a duration of about $\Delta t\sim r_g/c$ and a luminosity that approaches the maximum allowed power, $L_\gamma \sim \chi L_{BZ}$, where $\chi\simlt 1$ depends on the initial gap width (see Eq. (\ref{L_flare}) and text below),  followed by rapid, small amplitude oscillations that last for the entire simulation time, during which the pair plasma is continuously replenished through self-sustained pair creation bursts. The long-term behavior of the system (after the decay of the prompt spark) is essentially independent of the initial condition, provided that the initial pair density is well below the GJ density inside the gap. This behavior is reminiscent of the longitudinal oscillations found in the semi-analytic, two-beam model of \citet{2005ApJ...631..456L}.

\begin{figure*}[h]
\centering
\includegraphics[width=17.5cm]{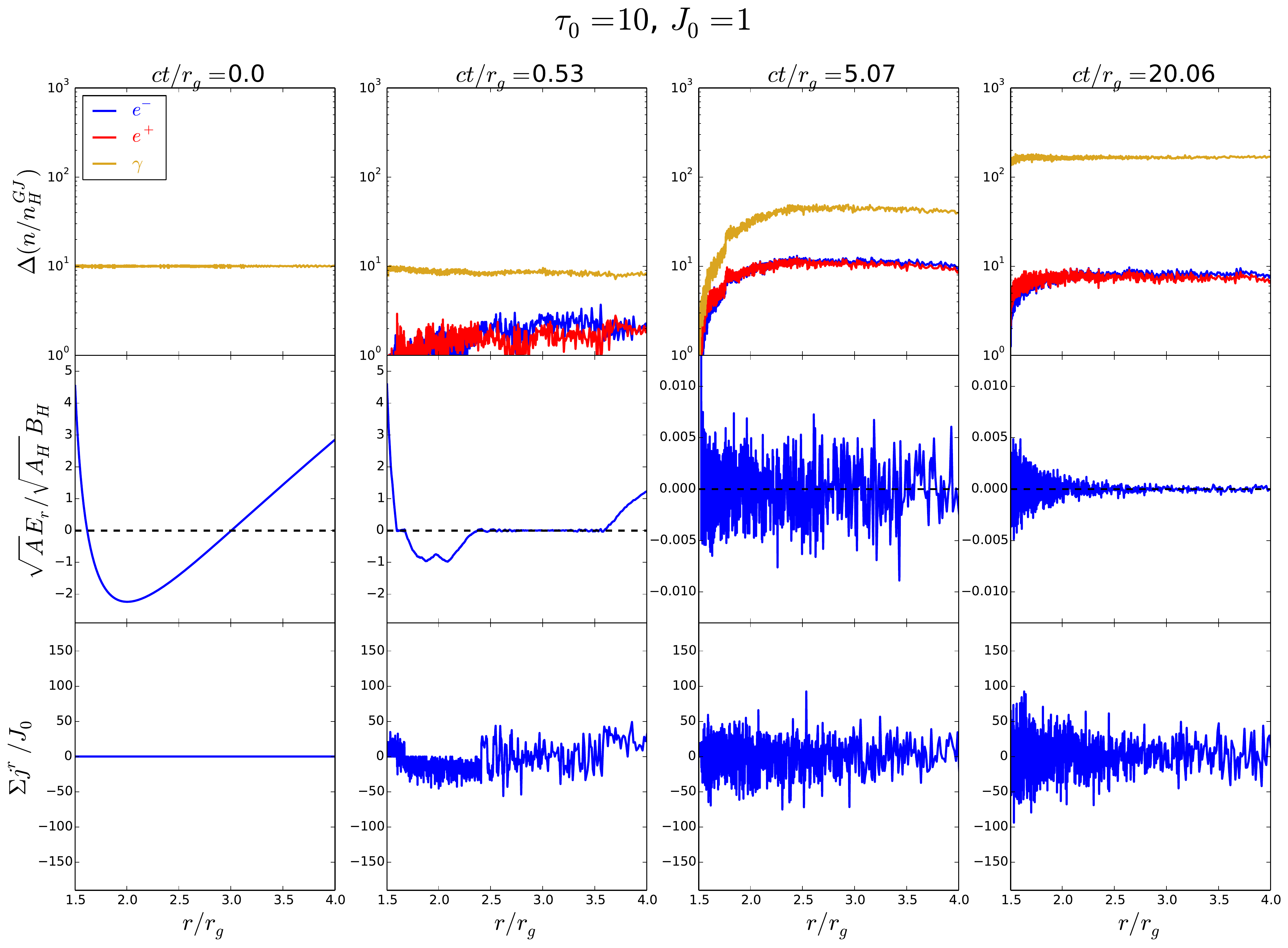}
\caption{Snapshots from a typical simulation of spark gap dynamics. Shown are the pair and photon densities (upper panels), electric flux (middle panels), and radial electric current, $\Sigma j^r$, normalized  by the global magnetospheric current $J_0$. The leftmost panels delineate the initial state, at $t=0$. The rightmost panels show the relaxed state, following the prompt discharge. We note the scale  change  on the vertical axis in the middle panels.}   
\label{fig:ex1}
\end{figure*}

\begin{figure*}[h]
\centering
\includegraphics[width=17.5cm]{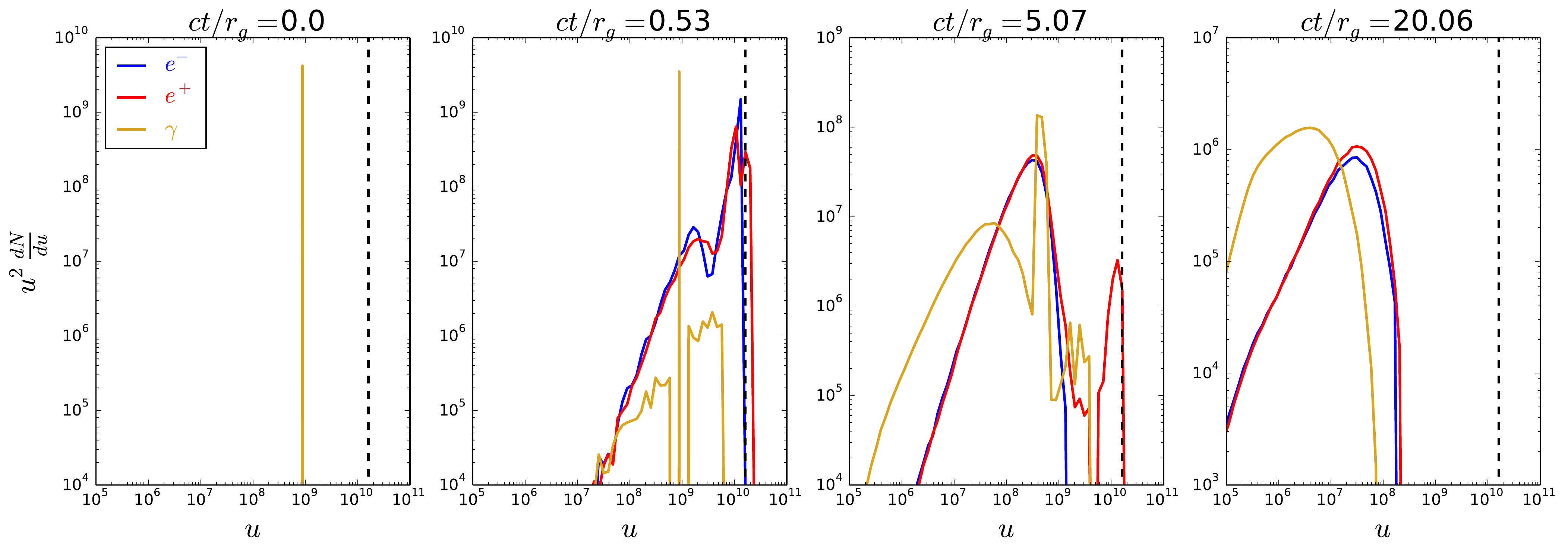}
\caption{Snapshots of electron (blue line), positron (red line), and photon (yellow line) spectra computed during the run presented in figure \ref{fig:ex1}.}   
\label{fig:ex1b}
\end{figure*}

A typical example is shown in Figs.  \ref{fig:ex1} and \ref{fig:ex1b}. Figure \ref{fig:ex1} exhibits four snapshots of the pair and photon densities (upper panels), electric flux (middle panels), and radial electric current (lower panels), and Fig.  \ref{fig:ex1b} the corresponding spectral energy distributions of pairs and photons. The initial pair density was taken to be zero in this example. To ignite the discharge process a mono-energetic gamma-ray distribution (left panel of Fig. \ref{fig:ex1b}) was injected at time $t=0$ inside the simulation box, with a uniform distribution in $\xi$ space. The initial energy of injected photons was chosen to optimize pair creation in order to speed up the computations. We also performed several runs with the same setup but different initial conditions, e.g., sub-GJ pair density and no photons, and verified that the overall evolution of the system is independent of the initial condition. In all cases, we find that after several crossing times the system approaches a state of quasi-steady oscillations  with essentially the same  properties (as in the rightmost panels in figures \ref{fig:ex1} and \ref{fig:ex1b}).

As seen in figure \ref{fig:ex1}, after about one crossing time from the start of the simulation the pair density exceeds the GJ value, giving rise to a nearly complete screening of the electric field.  The sporadic pair creation leads to rapid spatial and temporal oscillations of the electric field and radial current inside the simulations box. The pair density and the amplitude of the electric field oscillations ultimately approach their saturation levels (after several crossing times), at which time pair creation inside the simulation box balances pair escape through its boundaries. At this stage the pair and photon spectra are quasi-stationary.

\begin{figure}[h]
\centering
\includegraphics[width=8cm]{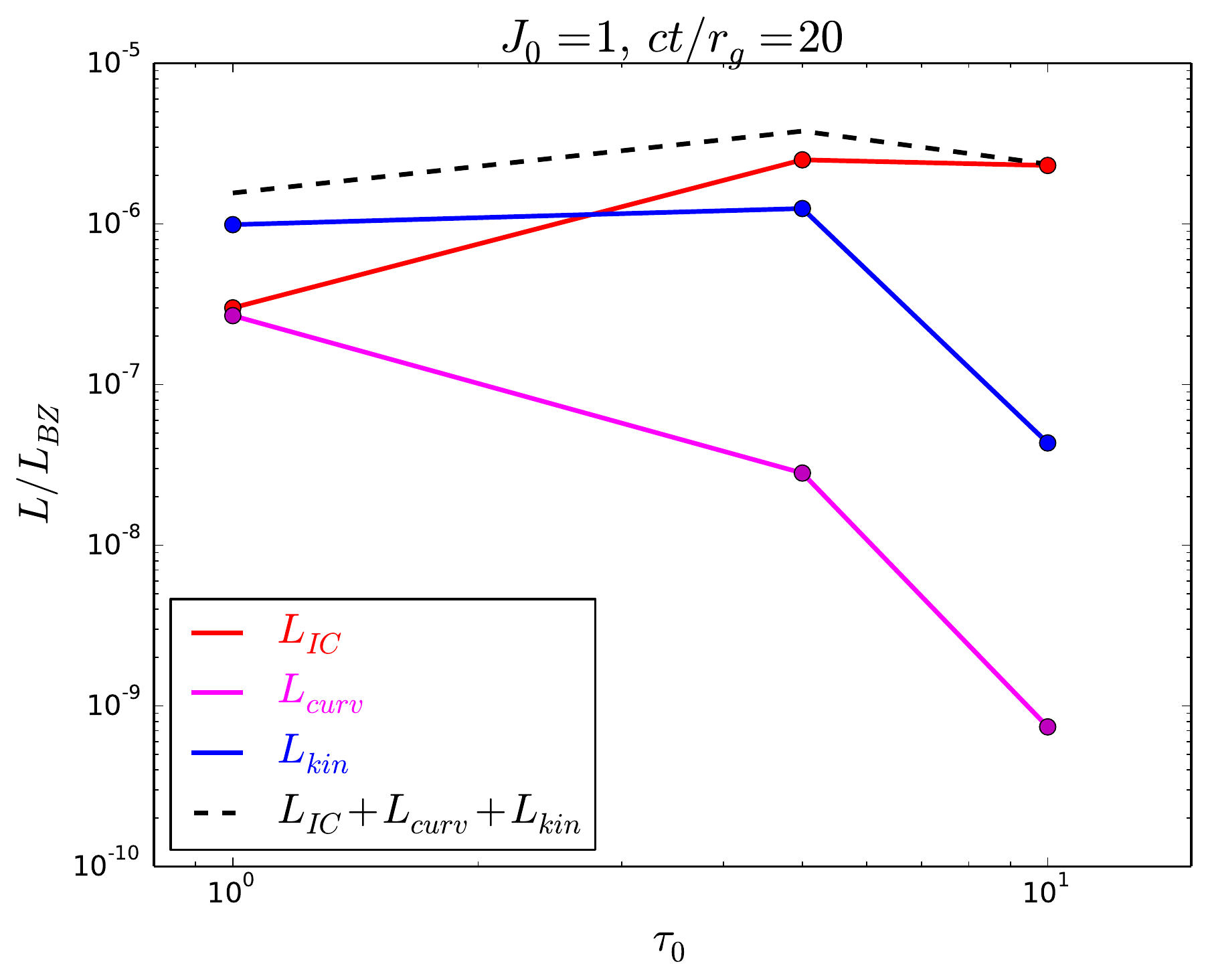}
\caption{Dependence of quiescent IC, curvature, and kinetic luminosities leaving the outer boundary of the box ($r_{\rm out}\approx 4$) on the fiducial opacity $\tau_0$.}   
\label{fig:Lgamma}
\end{figure}

Figure \ref{fig:j0} depicts the dependence of the pair multiplicity (upper panels), average Lorentz factor (middle panels) and average gamma-ray energy (lower panels) on the fiducial opacity $\tau_0$ (left) and global current $J_0$ (right). As can be seen, the multiplicity increases 
roughly linearly with $\tau_0$, while the average pair and photon energies decrease with increasing $\tau_0$. On the other hand, those quantities are essentially independent of $J_0$.    The overall gamma-ray luminosity emitted from the gap following the prompt phase depends weakly on $\tau_0$ (figure \ref{fig:Lgamma}). For the BZ power used to normalize the luminosities exhibited in figures \ref{fig:Lgamma} and \ref{fig:lightcurve} we adopt 
\begin{equation}
L_{BZ}=\frac{1}{16}\tilde{a}^2 B_H^2r_H^2.
\label{L_BZ}
\end{equation}
This is close to values obtained from solutions to the Grad--Shafranov equation \citep{2014ApJ...788..186N, 2018arXiv180200815M}, but may somewhat overestimate the values expected in realistic jets.

\subsection{Flaring states}

\begin{figure}[h]
\centering
\includegraphics[width=8cm]{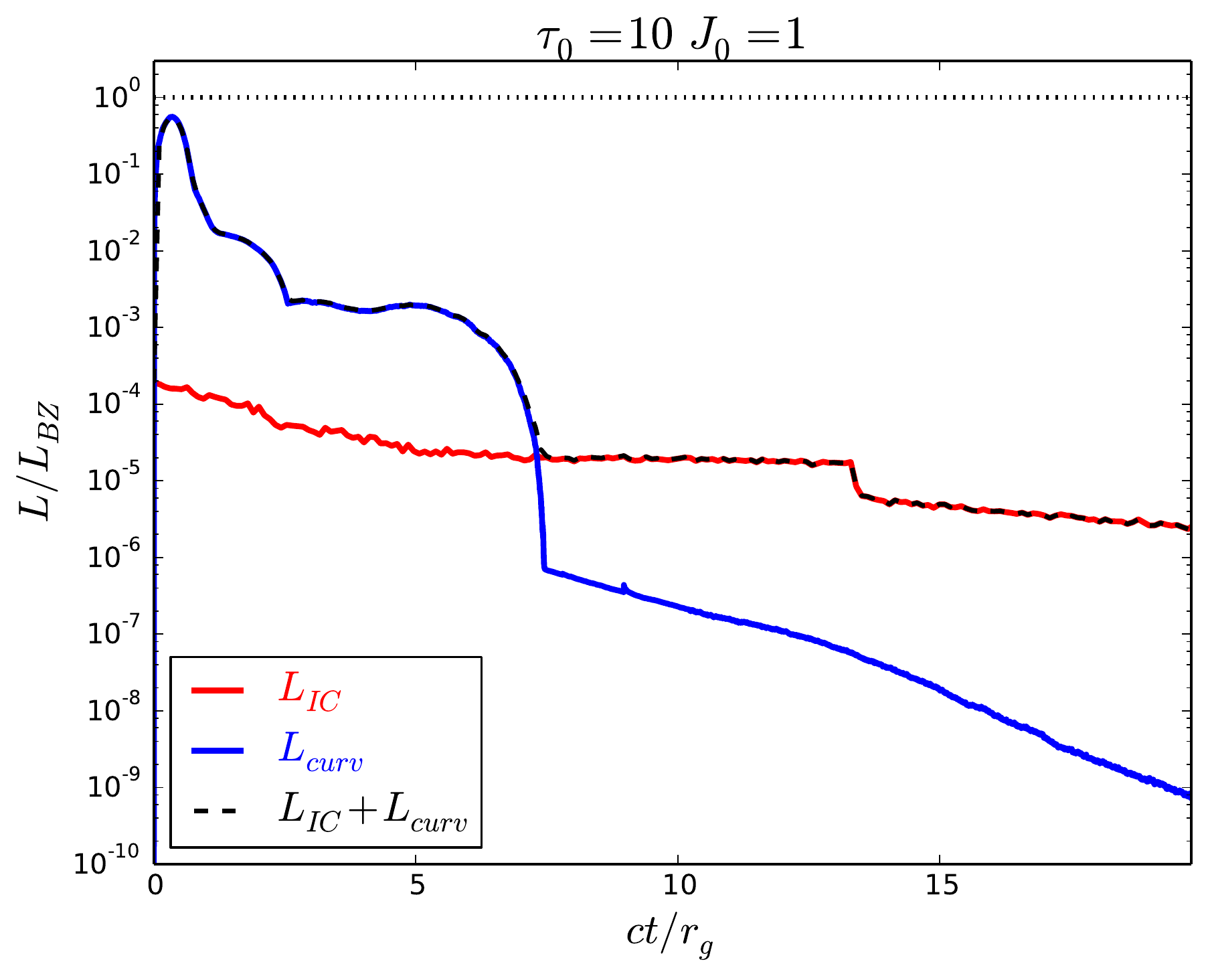}
\caption{Gamma-ray light curve produced by the gap discharge. The red line corresponds to IC emission, the blue line to curvature emission, and the black dashed line to the sum of both components.}   
\label{fig:lightcurve}
\end{figure}

During the initial discharge (at $t<r_g/c$), when the gap electric field is still sufficiently intense, the newly created pairs quickly accelerate to the terminal Lorentz factor $\gamma_{max}$ (Eq. (\ref{g_max})), at which time the energy gain is balanced by curvature losses (indicated by the vertical dashed line in figure \ref{fig:ex1b}). We note that at these energies Compton scattering is in the deep Klein--Nishina regime and is highly suppressed. As time passes the average pair energy declines, ultimately approaching a final value. For the case shown in figure \ref{fig:ex1b} it is about $5\times10^7$. At this state curvature emission is completely negligible (see Fig. \ref{fig:lightcurve}). As is evident from figure \ref{fig:ex1b}, during the initial discharge most particles quickly accelerate to the terminal Lorentz factor, $\gamma_i=\gamma_{max}$, where $P_{cur}(\gamma_i)\simeq e E$. The peak luminosity occurs roughly when the pair multiplicity approaches unity, while the electric field is not yet significantly screened out. Thus, the curvature luminosity can be approximated as $L_{cur}\simeq P
 _{cur} n_{GJ} 4\pi r_H^3/3 \simeq e E n_{GJ} 4\pi r_H^3/3$. From Equation (\ref{Poiss}) we estimate that $E\simeq e n_{GJ} r_H$ before screening ensues. With $e n_{GJ}\simeq \Omega B_H/2\pi $ and $\Omega=\omega_H/2=\tilde{a}/4r_H$, we obtain 
\begin{equation}
L_{cur}\simeq \frac{1}{12\pi} \tilde{a}^2B_H^2 r_H^2 \simeq L_{BZ},
\label{L_flare}
\end{equation}
where $L_{BZ}$ is the corresponding Blandford--Znajeck power given explicitly in Eq. (\ref{L_BZ}). Figure \ref{fig:lightcurve} exhibits the light curve computed from the PIC simulations.  As seen, the gamma-ray  luminosity indeed approaches the BZ power during the initial spark, and then decays to the terminal value as the gap electric field is screened out. 
It can be readily shown that if the initial gap width is small, $h \ll r_H$, then $L_{cur}$ is reduced by a factor $\chi\simeq (h/r_H)^2$.  

The above considerations suggest that strong rapid flares should be produced every time a magnetospheric gap is restored, for example by an accretion episode. The flaring episode is then followed by quiescent emission with a luminosity  $L_{\gamma}\sim 10^{-5} L_{BZ}$.

\section{\label{sec:M87}Applications to M87}

M87 exhibits TeV emission with a luminosity of $L_{TeV}\sim10^{40}$ erg s$^{-1}$ in the quiescent state. Several strong flares with durations of $\Delta t\simeq t_g$ have been 
recorded in the past decade \citep{2006Sci...314.1424A, 2008ApJ...685L..23A, 2009Sci...325..444A, 2012ApJ...746..151A}. 
Various estimates of the average jet power in M87 (see, e.g., \citealt{1996ApJ...467..597B, 2000ApJ...543..611O, 2006MNRAS.370..981S, 2009ApJ...699.1274B}) yield a range of a few times $10^{43}$ to a few times $10^{44}$ erg s$^{-1}$. Assuming that the jet is powered by the BZ mechanism implies $L_{TeV}/L_{BZ}\sim 10^{-4} $.
The SED exhibits a peak in the sub-mm band at $\nu_{peak}\simeq10^{12}$ Hz, corresponding to $\epsilon_{max}\simeq10^{-8}$ in our parametrization, with a bolometric luminosity of $L_b\simlt10^{41}$ erg s$^{-1}$. This sets an upper limit on the luminosity of the putative RIAF emission, $L_{s}=\eta_d L_b$, as the source of the observed SED is yet unresolved. For a BH mass of $M=6\times10^9 M_\odot$ this implies $l_s<10^{-7}$ and $\tau_0=10^3  \eta_d (\tilde{R}_s/10)^{-2}<10^3$ from Equation (\ref{tau_0}).
If the observed TeV emission originates from a spark gap, then the pair production opacity must be sufficiently low to allow TeV photons to escape the system.  For the target photon spectrum  invoked in Equation (\ref{Intensity-1}) with $p=2$, $\epsilon_{min}=10^{-8}$, the pair production optical depth is given approximately by $\tau_{\gamma\gamma}\simeq 0.1\tau_0 \tilde{R}_s(\epsilon_{max}\epsilon_{\gamma})^2 =10^{-13} \eta_d (\tilde{R_s}/10)^{-1} \epsilon_{\gamma}^2 $.  Thus, it is transparent at energies below about 3 TeV. However, 
the observed spectrum appears to extend up to $\sim 10$ TeV, implying $\eta_d \simlt10^{-1} $ and $\tau_0<10^2$. 

Speculating that the observed TeV emission is produced in a spark gap, we note that the gamma-ray power released by the gap, as predicted by our model, $L_\gamma\sim10^{-5}L_{BZ}$, is somewhat lower, but still consistent with the observed emission in the quiescent state given the various model uncertainties. Figure \ref{fig:ex1b} confirms that for $\tau_0=10$ the spectrum extends up to about 10 TeV. The strong flares recorded in the past two decades might be caused by some episodes during which the gap is suddenly restored. We note that a small percent of the full gap width is sufficient to account for the strongest flares observed by a sudden discharge.  Alternative models for the M87 flares include misaligned mini jets \citep{2010MNRAS.402.1649G}, and star--jet interactions (\citealt{2010ApJ...724.1517B}, see also \citealt{2017ApJ...841...61A}). 

\section{Conclusions}

We explored the dynamics of pair discharges in a starved magnetosphere of a Kerr BH, using 1D PIC simulations. Our analysis takes into account inverse Compton scattering and pair creation via the interaction of pairs and gamma rays with an ambient radiation field, assumed to be emitted by the putative accretion flow, as well as curvature losses. In the computations presented here the intensity of the external radiation field is taken to be a power law with a slope of $-2$ and  a minimum energy of $10^{-8}m_ec^2$ (corresponding to a minimum frequency of about $10^{12}$ Hz).
We find, quite generally, that the initial gap electric field is screened out by a prompt discharge of duration $\sim r_g/c$ that produces a strong flare of very high-energy curvature photons. This episode is followed by a state of self-sustained, rapid plasma oscillations that lasts for the entire simulation time, during which the pair and gamma-ray spectra are quasi-stationary, with a gamma-ray luminosity that constitutes a fraction of about $10^{-5}$ of the BZ power depending weakly on input parameters. As pointed out in \S \ref{sec:M87}, this value is in rough agreement with observations of M87.

It is worth noting that the ratio $L_\gamma/L_{BZ}$ obtained during the quiescent state in our model is typically smaller than the values obtained in steady-state models with extremely low accretion  rates. For example, \citet{2017ApJ...845...77H} obtained values in the range $10^{-4} - 10^{-1}$ for accretion rates in the range from $10^{-4}$ to  $5\times10^{-6}$ Eddington, assuming an equipartition magnetic field in the disk. Thus, if the gaps are unsteady, as our simulations seem to indicate, reevaluation of observational predictions may be needed.
On the other hand, for the same accretion rates the BZ power can be larger by up to an order of magnitude if the magnetic field approaches the saturation level predicted by MAD models, as seems to be suggested by observations of M87. This will give rise to correspondingly higher gamma-ray luminosities.

Our choice of parameters in the present analysis was motivated by the applications to M87 and conceivably other AGNs. However, the gap emission may also be relevant  to galactic BH transients \citep{2017ApJ...845...40L, 2017PhRvD..96l3006L}, which span a different regime in parameter space. If similar $L_\gamma/L_{BZ}$ ratios are produced by Galactic BHs, then a 10 solar
mass BH accreting at a rate of $10^{-3}$ Eddington, can conceivably produce a TeV luminosity of about $10^{31}$ erg s$^{-1}$ that can be detected by the Cherenkov Telescope Array out to a distance of about 1 kpc. However, it is plausible that in these sources the emission will be dominated by curvature losses due to the smaller curvature radius, and it remains to be seen how the shape of the emitted spectrum scales with source parameters. We leave this problem to a future work.

Ultimately, global PIC simulations are required to compute the full structure of the magnetosphere and its response to pair discharges in starved region. Moreover, our 1D model is restricted to longitudinal plasma oscillations, while in reality transverse modes might be excited by the gap activity, which can only be accounted for in 2D simulations. Nonetheless, our model captures the basic features of plasma production and gap emission.

\begin{acknowledgements}
We thank Maxim Barkov and Alexander Philippov for their comments on the manuscript.
AL acknowledges the kind hospitality of IPAG, where the essential part of this work was done, and the support from the visiting professor program of the Universit\'e Grenoble Alpes.   
BC acknowledges support from CNES and Labex OSUG@2020 (ANR10 LABX56). This work was granted access to the HPC resources of TGCC under the allocation t2016047669 made by GENCI.  
\end{acknowledgements} 

\onecolumn
\begin{appendix}
\section{\label{sec:EM} Derivation of the gap electrodynamic equations}

We restrict our analysis to axisymmetric systems. We suppose that outside the gap the flow is stationary, and that inside the gap particles can only move along magnetic surfaces; i.e., the gap oscillations are longitudinal (electrostatic). In the force-free section outside the gap the angular velocity of field lines, $\Omega$, is fixed. If the gap forms a small perturbation in the magnetosphere, then variation in $\Omega$ across the  gap can be ignored.   It is then appropriate to define the electric field in the corotating frame as $F^\prime_{\mu t}=F_{\mu t}+\Omega F_{\mu \varphi}$. Outside the gap, in the ideal MHD section, $F^\prime_{\mu t}=0$. Inside the gap, Gauss's law, $\partial_\mu (\sqrt{-g} F^{t\mu})=\sqrt{-g}j^t$, reduces to (see, e.g., \citealt{2017PhRvD..96l3006L})
\begin{equation}
\left[\frac{\Delta \sin\theta}{\alpha^2} F^\prime_{rt}\right]_{,r}
+\left[\frac{\sin\theta}{\alpha^2} F^\prime_{\theta t}\right]_{,\theta}
=4\pi\sqrt{-g}(j^{t}-\rho_{GJ}),\label{Max_t_1D}\\
\end{equation}
where the GJ density is given by
\begin{equation}
4\pi \sqrt{-g}\rho_{GJ}=\left[\frac{\Delta \sin\theta}{\alpha^2}(\omega-\Omega) F_{r \varphi}\right]_{,r}
+\left[\frac{\sin\theta}{\alpha^2}(\omega-\Omega) F_{\theta \phi}\right]_{,\theta}
= A_0\left[\frac{\sin^2\theta}{\alpha^2}(\omega-\Omega)\right]_{,\theta},\label{rho_GJ_1D}
\end{equation}
and the last equality applies to the split monopole geometry invoked in our model, $A_\varphi=C(1-\cos\theta)$.  
We note that $\sqrt{A} B_r \sin\theta = F_{\theta\varphi}=C \sin\theta$, which implies that $\sqrt{A} B_r=\sqrt{A_H} B_H= C$, here
$B_H=B_r(r=r_H)$ and likewise $A$. Explicitly:
\begin{equation}
\rho_{GJ} =\frac{B_H\sqrt{A_H}\cos\theta}{2\pi \Sigma^2 \Delta }\left[\left(A+\frac{2Mr(r^2+a^2)}{\Sigma}a^2\sin^2\theta \right)(\omega-\Omega)+\Delta\,\omega\,a^2\sin^2\theta\right].
\label{app:rho_GJ_exp}
\end{equation}
Since the gap dynamics is restricted to longitudinal oscillations we have $F^\prime_{\theta t}=0$ in Eq. (\ref{Max_t_1D}). Next, the radial component of Amp\`ere's law, $\partial_\mu (\sqrt{-g} F^{r\mu})=\sqrt{-g}j^r$, gives
\begin{equation}
-\left[\frac{\Delta \sin\theta}{\alpha^2}\{F^\prime_{rt}+(\omega-\Omega) F_{r \varphi}\}\right]_{,t}
+\left(\frac{\Delta\sin\theta}{\Sigma}F_{r\theta}\right)_{,\theta}=4\pi\sqrt{-g}j^r,\label{Amp_r}
\end{equation}
and for our split monopole geometry $F_{r\varphi}=0$.  Outside the gap the flow is stationary ($\partial_t=0$), and charge conservation, $\partial_\mu(\sqrt{-g} j^\mu)=0$, yields $\partial_r(\Sigma j^r)=0$ in the force-free limit, so that the radial electric current is conserved along magnetic surfaces: $\Sigma j^r=J_0=$ const. From Equation (\ref{Amp_r}) we obtain 
\begin{equation}
J_0=\frac{1}{4\pi \sin\theta}\left(\frac{\Delta\sin\theta}{\Sigma}F_{r\theta}\right)_{,\theta}\label{J_0}
\end{equation}
outside the gap.  This conserved current must be flowing through the gap. Thus, the induction 
equation, Eq. (\ref{Amp_r}), reduces to 
\begin{equation}
\partial_t \left(\frac{A}{\Sigma}F^\prime_{rt}\right) =-4\pi(\Sigma j^r-J_0).
\label{app:dtEr}
\end{equation}
Frame dragging couples the toroidal and poloidal components of the wave field:
\begin{eqnarray}
\partial_t \left(\frac{\Sigma}{\sqrt{\Delta}}E^\prime_{\varphi}\right) &=& \sqrt{A}\sin\theta E^\prime_r \partial_r \omega,\\
\partial_t \left(\frac{\Sigma}{\sqrt{\Delta}}B^\prime_{\varphi}\right) &=& \partial_\theta \left(\frac{\Sigma}{\sqrt{A}}E_r^\prime\right).
\end{eqnarray}
From the latter relations we find $|E^\prime_\varphi|\sim (\lambda/r)|E^\prime_r |\ll|E^\prime_\varphi|$, where $\lambda \ll r$ is the characteristic wavelength of the oscillations, and likewise for $B_\varphi^\prime$,  thus those fields can be neglected.

\section{\label{app:MC}Monte Carlo scheme}
\subsection{Compton scattering}
The Compton scattering opacity is given by
\begin{eqnarray}\label{k_c}
\kappa_c(\gamma)&=&\frac{2\pi\sigma_T}{h c}\int_{-1}^1d\mu_s(1-\beta\mu_s) \int_{\epsilon_{min}}^{\epsilon_{max}} \frac{d\epsilon_s}{\epsilon_s} I_s(\epsilon_s)\tilde{\sigma}_{KN}(\epsilon^\prime_s)
=\frac{\tau_0}{2 r_g}\int_{-1}^1d\mu_s(1-\beta\mu_s) \int_{\epsilon_{min}}^{\epsilon_{max}} \frac{\epsilon_{min}^pd\epsilon_s}{\epsilon_s^{p+1}}\tilde{\sigma}_{KN}(\epsilon^\prime_s),
\end{eqnarray}
where $\gamma$ is the Lorentz factor of the particle before scattering, $I_s(\epsilon_s)$ is the intensity of the ambient radiation field defined in Equation (\ref{Intensity-1}), $\epsilon^\prime_s=\gamma(1-\beta\mu_s)\epsilon_s $ is the target photon energy in the rest frame of the particle, $\tau_0$ is the fiducial optical depth given in Equation (\ref{tau_0}), and 
\begin{eqnarray}
\tilde{\sigma}_{\rm KN}(x) = \frac{3}{4}
 \left[ \frac{1 + x}{x^3}
          \left\{ \frac{2 x (1+x)}{1+2 x} - {\rm ln}(1+2x) \right\}
 + \frac{1}{2x} {\rm ln}(1+2 x) - \frac{1 + 3x}{(1+2x)^2}
   \right]
\end{eqnarray}
is the Klein--Nishina cross section measured in units of $\sigma_T$. The opacity $\kappa_c(\gamma)$ is used to draw scattering events.

\subsubsection*{\it Drawing the scattered photon energy}
Once a scattering event occurs, the energy of the scattered  photon is drawn upon transforming to the rest frame of the scatterer. 
The  probability density that an electron (positron) will produce a gamma ray of rest frame energy $\epsilon_\gamma^\prime=\gamma(1-\beta\mu_\gamma)\epsilon_\gamma$ in a single scattering can be expressed as
\begin{equation}
f(\epsilon_\gamma^\prime)=A\int_{\epsilon^\prime_\gamma}^{\epsilon^\prime_u}d\epsilon^\prime_s\int_{\mu^\prime_{min}}^{\mu^\prime_{max}}d\mu_s^\prime \sigma(\epsilon^\prime_\gamma,\epsilon^\prime_s)n_s^\prime(\epsilon^\prime_s,\mu^\prime_s),
\label{comp_feg}
\end{equation}
where $A$ is a normalization coefficient, defined such that $\int f(\epsilon_\gamma^\prime)d\esp=1$,
$\epsilon^\prime_u=$min\{$\epsilon_\gamma^\prime/(1-2\epsilon_\gamma^\prime)$,$\epsilon_{max}^\prime$\},
$\beta\mu_{min}^\prime=$ max $\{-\beta, (\epsilon_{min}/\gamma\epsilon^\prime_s)-1\}$, with $-\beta$ holding for 
$\epsilon^\prime_s>\epsilon_{min}/\gamma(1-\beta)$, 
and $\beta\mu_{max}^\prime=$ min $\{\beta, (\epsilon_{max}/\gamma\epsilon^\prime_s)-1\}$, with $\beta$  holding for 
$\epsilon^\prime_s < \epsilon_{max}/\gamma(1+\beta)$, and
\begin{equation}
n_s^\prime(\epsilon^\prime_s,\mu^\prime_s)d\epsilon_s^\prime=\frac{I_s[\epsilon_s^\prime \gamma(1+\beta\mu_s^\prime]}{[\gamma(1+\beta\mu_s^\prime)]^3}\frac{d\epsilon_s^\prime}{\epsilon_s^\prime},\quad \gamma(1-\beta)\epsilon_{min}<\epsilon^\prime_s<\gamma(1+\beta)\epsilon_{max}
\label{Spect_n}
\end{equation}
is the spectral density of target radiation field in the rest frame of the particle.
The cross section for scattering of a photon of energy $\epsilon^\prime_s$ and direction $\mu^\prime_s$ to a final energy $\epsilon^\prime_\gamma$ is given explicitly by
\begin{equation}
\sigma(\epsilon^\prime_\gamma,\epsilon^\prime_s)=\int \frac{d\sigma}{d\Omega}\delta\left[\epsilon_\gamma^\prime-\frac{\epsilon_s^\prime}{1+\epsilon_s^\prime(1-\cos\Theta)} \right]d\varphi_\gamma^\prime d\mu_\gamma^\prime
=\frac{2\pi}{\epsilon^{\prime 2}_\gamma}\frac{d\sigma}{d\Omega}(\epsilon_s^\prime,\epsilon_\gamma^\prime),
\end{equation}
in terms of the differential cross-section
\begin{equation}
\frac{d\sigma}{d\Omega}=\frac{3\sigma_T}{16\pi}\left(\frac{\epsilon_\gamma^\prime}{\epsilon_s^\prime}\right)^2
\left(\frac{\epsilon_s^\prime}{\epsilon_\gamma^\prime}+\frac{\epsilon_\gamma^\prime}{\epsilon_s^\prime}-\sin^2\Theta \right)
\label{Diff_KN}
,\end{equation}
where $\cos\Theta=\cos\theta^\prime_s\cos\theta^\prime_\gamma-\sin\theta^\prime_s\sin\theta^\prime_\gamma\cos(\varphi_s^\prime-\varphi^\prime_\gamma)$ is the angle between incident and scattered photons in the particle's rest frame, and $d\sigma(\esp,\egp)/d\Omega$  is obtained by substituting $\cos\Theta=1+1/\epsilon_s^\prime -1/\epsilon_\gamma^\prime$ in Equation (\ref{Diff_KN}).

To save computing time we use an approximate method to sample the probability density $f(\egp)$. We first note that to a good approximation $\sigma(\epsilon^\prime_\gamma,\epsilon^\prime_s)\simeq \frac{3 }{4 \epsilon_\gamma^{\prime }\esp}$. Then, substituting Equations (\ref{Intensity-1}), (\ref{Spect_n}), and the latter relation into Equation  (\ref{comp_feg}), we obtain 
\begin{equation}
f(\epsilon_\gamma^\prime)\simeq \frac{3  }{4 \epsilon_\gamma^{\prime }}\int_{\epsilon^\prime_\gamma}^{\epsilon_u^\prime} 
d\epsilon^\prime_s \epsilon^{\prime -(p+2)}_s 
 \int_{\mu^\prime_{min}}^{\mu^\prime_{max}}d\mu_s^\prime [\gamma(1+\beta\mu_s^\prime)]^{-(3+p)}.
\end{equation}
The cumulative distribution is given by
\begin{equation}
F(\epsilon_\gamma^\prime)=\frac{\int_{\frac{\eminp}{1+2\eminp}}^{\epsilon^\prime_\gamma}  f(\epsilon_\gamma^\prime) d\epsilon^\prime_\gamma }
{\int_{\frac{\eminp}{1+2\eminp}}^{\epsilon^\prime_{max}}  f(\epsilon_\gamma^\prime) d\epsilon^\prime_\gamma}.
\end{equation}
To shorten the notation we denote $\epsilon_r=\gamma(1+\beta)\epsilon_{min}$.
Approximate analytic expressions for $F$ are obtained in terms of $\epsilon_r$ as follows:

For $\epsilon_r\le 0.3$,
\begin{equation}\label{E_dist_cumul}
F(\epsilon_\gamma^\prime)=C_1^{-1}
\left\{
 \begin{array}{lr}
   ( \epsilon_\gamma^{\prime}/\epsilon_r)^2-(\epsilon^{\prime}_{\gamma min}/\epsilon_r)^2& : \egp < \epsilon_{r}\\
  1+2/p - (2/p)( \epsilon_r/\epsilon_\gamma^{\prime})^p  -(\epsilon^{\prime}_{\gamma min}/\epsilon_r)^2  & : \egp > \epsilon_{r}
  \end{array}
  \right.
\end{equation}
where 
$
C_1=1+2/p - (2/p)( \epsilon_r/\epsilon_{\gamma max}^{\prime})^p  -(\epsilon^{\prime}_{\gamma min}/\epsilon_r)^2.
$

\begin{figure}[h]
\centering
\includegraphics[width=12cm]{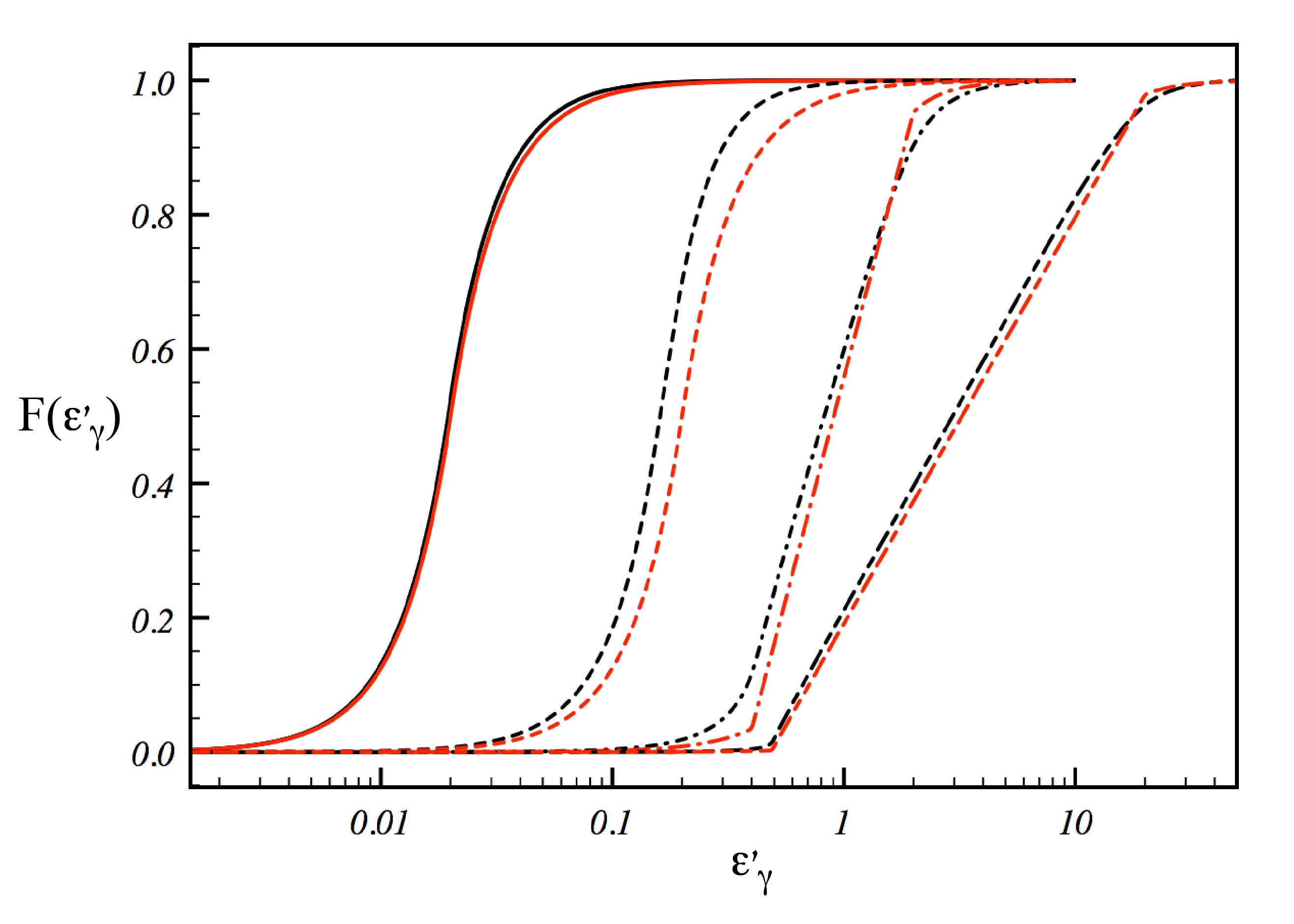}
\caption{Cumulative probability distribution for selecting scattered photon energy, for different values of $\epsilon_r=\gamma(1+\beta)\epsilon_{min}$, with $\epsilon_r=0.02$ (solid lines), $0.2$ (dashed lines), 2 (dot-dashed lines), 
and 20 (long-short dashed lines). The black curves delineate the exact distribution, and the red curves the analytic approximation, Equations (\ref{E_dist_cumul}) and (\ref{E_dist_cumul_KN}).}   
\label{fig:E_draw1}
\end{figure}

For $\epsilon_r > 0.3$,
\begin{equation}\label{E_dist_cumul_KN}
F(\epsilon_\gamma^\prime)=C_2^{-1}
\left\{
 \begin{array}{lr}
   ( \epsilon_\gamma^{\prime}/\epsilon_r)^2-(\epsilon^{\prime}_{\gamma min}/\epsilon_r)^2& : \egp < \epsilon_{r}/(1+2\epsilon_r)\\
  \frac{p+2}{p+1}\epsilon_r\ln(\egp/\epsilon_r+2\egp) + 1/(1+2\epsilon_r)^2 -(\epsilon^{\prime}_{\gamma min}/\epsilon_r)^2& : \epsilon_r>\egp > \epsilon_{r}/(1+2\epsilon_r)\\
 - (\epsilon_r/\egp)^{p+1}\epsilon_r/(p+1)^2 + G_0 &  \egp >\epsilon_r
  \end{array}
  \right.
\end{equation}
where 
$C_2= \epsilon_r/(1+p)^2 + \frac{p+2}{p+1}\epsilon_r\ln(1+2\epsilon_r) + 1/(1+2\epsilon_r)^2 -(\epsilon^{\prime}_{\gamma min}/\epsilon_r)^2$.
The analytic distribution, Equations (\ref{E_dist_cumul}) and  (\ref{E_dist_cumul_KN}), is invertible and can be readily used to randomly select the energy $\egp$. It is plotted in Fig. \ref{fig:E_draw1} (red lines) for different values of $\epsilon_r=\gamma(1+\beta)\epsilon_{min}$, and compared (black lines) with the exact expression computed by numerically integrating Equation (\ref{comp_feg}). 

\subsubsection*{\it Drawing the scattered photon direction}
Once the energy $\egp$  is selected it can be used to draw the direction $\mu_\gamma^\prime$ of the scattered photon.
The cross section for a scattering of target photon of energy $\esp$ into the final state $\egp$, $\mu_\gamma^\prime$ is
\begin{eqnarray}
\sigma(\epsilon^\prime_\gamma,\epsilon^\prime_s,\mu_\gamma^\prime)
=\int \frac{d\sigma}{d\Omega}\delta\left[\epsilon_\gamma^\prime-\frac{\epsilon_s^\prime}{1+\epsilon_s^\prime(1-\cos\Theta)} \right]d\varphi_\gamma^\prime 
=\frac{1}{\epsilon^{\prime 2}_\gamma\sqrt{(1-\chi^2)(1-\mu_\gamma^{\prime2})-(\mu_s^\prime-\mu_\gamma^{\prime}\chi)^2}}\frac{d\sigma}{d\Omega}(\epsilon_s^\prime,\epsilon_\gamma^\prime),
\end{eqnarray}
where 
\begin{equation}
\chi(\esp,\egp)=1+\frac{1}{\epsilon_s^\prime}-\frac{1}{\epsilon_\gamma^\prime}.
\label{scatt-angle}
\end{equation}
The conditional probability distribution reads 
\begin{equation}
f(\mu_\gamma^\prime|\epsilon_\gamma^\prime)=\frac{\int_{\epsilon^\prime_\gamma}^{\epsilon^\prime_u}d\epsilon^\prime_s\int_{\mu^\prime_{s-}}^{\mu_{s+}^\prime} d\mu_s^\prime \sigma(\epsilon^\prime_\gamma,\epsilon^\prime_s,\mu_\gamma^\prime)n_s^\prime(\epsilon^\prime_s,\mu^\prime_s)}
{f(\epsilon_\gamma^\prime)},
\end{equation}
with $\mu^\prime_{s\pm}=\mu^\prime_\gamma\chi\pm\sqrt{1-\mu^{\prime2}_\gamma}\sqrt{1-\chi^2}$, and $f(\egp)$ given by Equation (\ref{comp_feg}). Due to the extremely strong beaming we can safely assume that all particles arrive from direction $\mu_s^\prime=-1$ with a very small scatter. The latter relation then yields $\chi=-\mu_\gamma^\prime$, and from Eq. \ref{scatt-angle} we have
\begin{equation}
\frac{1}{\esp}=\frac{1}{\egp}-(1+\mu_\gamma^\prime); \quad -1\le \mu_\gamma^\prime\le {\rm min}\{1, 1/\egp -1 \}.
\end{equation}
Approximating the cross section by $\sigma(\epsilon^\prime_\gamma,\epsilon^\prime_s)\simeq \frac{3 }{4 \epsilon_\gamma^{\prime }\esp}$ and recalling that $n_s\propto \epsilon_s^{\prime -(p+1)}$, we find
\begin{equation}
f(\mu_\gamma^\prime|\epsilon_\gamma^\prime)\propto \left[\frac{1}{\egp}-(1+\mu_\gamma^\prime)\right]^{p+2}; \quad -1\le \mu_\gamma^\prime\le {\rm min}\{1, 1/\egp -1 \}.
\end{equation}
 The cumulative distribution, $F(\mu_\gamma^\prime|\egp)=F_0^{-1}\int_{-1}^{\mu_{\gamma}^\prime}f(y|\egp)dy$, explicitly given by
$$
F(\mu_\gamma^\prime|\epsilon_\gamma^\prime)
=\frac{1-\left[1-\egp (1+\mu_\gamma^\prime)\right]^{p+3}}{F_0}
 \quad -1\le \mu_\gamma^\prime\le {\rm min}\{1, 1/\egp -1 \},
$$
where $F_0=1-(1 -2\egp)^{p+3}$ if $\egp\le 1/2$ and $F_0=1$ if $\egp>1/2$, is used to randomly select $\mu^\prime_\gamma$.

\subsection{Pair production}
The full pair production cross section (measured in units of $\sigma_T$) is given by
\begin{equation}
\tilde{\sigma}_{\gamma\gamma}(\epsilon_s,\epsilon_\gamma,\mu)=\frac{3}{16}(1-\beta_{cm}^2)\left[(3-\beta_{cm}^4)\ln\left(\frac{1+\beta_{cm}}{1-\beta_{cm}}\right)-2\beta_{cm}(2-\beta_{cm})\right]
,\end{equation}
where $\beta_{cm}$ is the speed of the electron and positron in the center of momentum frame, given by
\begin{equation}
1-\beta_{cm}^2=\frac{2}{(1-\mu)\epsilon_s\epsilon_\gamma}, 
\end{equation}
and $\epsilon_\gamma$, $\epsilon_s$ are the dimensionless energies of the annihilating photons. 
The threshold energy for pair production through the interaction of a target photon with a gamma ray of energy $\epsilon_\gamma$ is given by the condition $\beta_{cm}=0$ or $\epsilon_{th}=2/(1-\mu)\epsilon_\gamma$.
The pair production opacity reads
\begin{eqnarray}\label{k_PP}
\kappa_{pp}(\epsilon_\gamma)&=&\frac{2\pi \sigma_T}{h c}\int_{-1}^1d\mu(1-\mu)\int_{\epsilon_{th}}^{\epsilon_{max}}\frac{d\epsilon_s}{\epsilon_s}I(\epsilon_s,\mu)\tilde{\sigma}_{\gamma\gamma}(\epsilon_s,\epsilon_\gamma,\mu)
=\frac{\tau_0}{2 r_g}\int_{-1}^1d\mu(1-\mu) \int_{\epsilon_{th}}^{\epsilon_{max}} \frac{\epsilon_{min}^pd\epsilon_s}{\epsilon_s^{p+1}}\tilde{\sigma}_{\gamma\gamma}(\epsilon^\prime_s).
\end{eqnarray}
\end{appendix}

\twocolumn

\end{document}